\documentclass[article]{JHEP3} 

\usepackage{epsfig,multicol,bbm}
\newcommand\fverb{\setbox\fverbbox=\hbox\bgroup\verb}
\newcommand\fverbdo{\egroup\medskip\noindent%
			\fbox{\unhbox\fverbbox}\ }
\newcommand\fverbit{\egroup\item[\fbox{\unhbox\fverbbox}]}
\newbox\fverbbox

\title{A unified picture of phase transition: from liquid-vapour systems to AdS black holes}

\author{Rabin Banerjee, Sujoy Kumar Modak and Dibakar Roychowdhury\\
	S.N. Bose National Centre for Basic Sciences\\ Block-JD, Sector-III, Salt Lake City, Kolkata-700098, India \\
	E-mail: \email{rabin@bose.res.in\\}E-mail: \email{sujoy@bose.res.in\\}Email: \email{dibakar@bose.res.in}}

\abstract{Based on fundamental concepts of thermodynamics we examine phase transitions in black holes defined in Anti-de Sitter (AdS) spaces. The method is in line with that used a long ago to understand the liquid-vapour phase transition where the first order derivatives of Gibbs potential are discontinuous and Clausius-Clapeyron equation is satisfied. The idea here is to consider the AdS black holes as grand-canonical ensembles and study phase transition defined by the discontinuity of second order derivatives of Gibbs potential. We analytically check that this phase transition between the `smaller' and `larger' mass black holes obey Ehrenfest relations defined at the critical point and hence confirm a second order phase transition. This include both the rotating and charged black holes in Einstein gravity.}


\begin{document} 


\section{Introduction}
It is well known that black holes behave as thermodynamic systems. The laws of black hole mechanics become similar to the usual laws of thermodynamics after appropriate identifications between the black hole parameters and the thermodynamical variables \cite{bch}. Nonetheless black holes show some exotic behaviors as a thermodynamic system, such as (i) entropy of black holes is proportional to area and not the volume, (ii) their temperature diverges to infinity when mass tends to vanish due to Hawking evaporation. In fact these exotic behaviors enhance the scope of alternative studies on black holes. The thermodynamic properties of black holes were further elaborated in early eighties when Hawking and Page discovered a phase transition between the Schwarzschild AdS black hole to thermal AdS space \cite{hp1}. Thereafter this subject has been intensely studied in various viewpoints \cite{hp2}-\cite{hp14}. 


The gibbsian approach to the phase transitions is based on the Clausius-Clapeyron-Ehrenfest's equations \cite{stanley,zeman}. These equations allow for a classification of phase transitions as first order or continuous (higher order) transitions. For a first order transition the first order derivatives (entropy and volume) of the Gibbs free energy is discontinuous and Clausius-Clapeyron equation is satisfied. Liquid to vapour transition is an ideal example of such a phase transition. Similarly for a second order transition Ehrenfest's relations are satisfied. However, this specific tool concerning the nature of phase transition has never been systematically used for black holes. Recently, in a series of papers \cite{bss}-\cite{bgr} we have developed an approach based on these concepts to study phase transitions in black holes. Despite of some conceptual issues mentioned in the earlier paragraph, we found that black holes not only allow us to implement above basic ideas but also support them. In normal physical systems a verification of the Clausius-Clapeyron or Ehrenfest relation is possible subjected to the design of a table top experiment. However black holes are found to be like a theorist's laboratory, where these ideas on phase transition could be proved without performing any table top experiment.

In our earlier work for black hole phase transition we did not notice any discontinuity in the first order derivative of Gibbs energy and hence Clausius-Clapeyron equation was not needed. However, bacause of the discontinuity in second order derivatives we derived Ehrenfest's relations for black holes in \cite{bss}. Because of infinite divergences of various physical entities at the critical point, we took recourse to numerical methods to check the validity of the Ehrenfest's relations. A similar analysis was performed in \cite{bgr} for the RN AdS black hole. In both cases we found that, away from the critical point, only the first Ehrenfest's relation was satisfied. We further developed these ideas in \cite{modak} and were able to verify both Ehrenfest's relations infinitesimally close to the critical point. This analysis was again based numerical computations.  Nevertheless it should be emphasised that such a numeric check has a drawback. Since in this method one assigns a numerical value for the angular velocity or eletric potential beforehand and then check the Ehrenfest relations, everytime this value is changed it is necessary to repeat the numerical analysis.  This add a limitation for generalising the result for other cases.

In this paper we develop an elegant analytical technique to treat the singularities near the critical points of the phase transition curves. This is important for further studies on black hole phase transition using our techniques. The example of the charged (RN) AdS and rotating (Kerr) black holes are worked out in details. We show analytically that both Ehrenfest's relations are satisfied exactly. This confirms the onset of a second order phase transition for these black holes. Explicit expressions for the critical temperature, critical mass etc. are given. This phase transition is characterized  by the divergence of the specific heat at the critical temperature. The sign of specific heat is different in two phases and they essentially separate two branches of AdS black holes with different mass/ horizon radius. The branch with lower mass (horizon radius) has a negative specific heat and thus falls in an unstable phase. The other branch with larger mass (horizon radius) is locally stable since it is associated with a positive specific heat and also positive Gibbs energy. Our earlier results for the Kerr AdS \cite{modak}, based on the numerical analysis, are also confirmed by the present analytical scheme. Thus the present paper complets the developement initiated in the earlier works  \cite{bss}-\cite{bgr}. {\footnote{As a point of caution we would like to mention that the phase transition studied in this paper should not be confused with the well known Hawking-Page (HP) phase transition\cite{wit1}. The latter is a transition between the AdS space and black hole phase \cite{hp9}-\cite{hp10}. As mentioned in earlier works \cite{hp7},\cite{hp10} such (HP) phase transition occurs at a temperature which is greater than the temperature where specific heat diverges. It is the latter case, which deals with a phase transition between two black hole phases, is considered in this paper.}} 


\section{Liquid to vapour phase transition} 
Liquid and vapour are just two phases of matter of a single constituent. When heated upto the boiling point liquids vapourise and at this point the slope of the coexistence curve (pressure $P$ versus temperature $T$ plot) obeys the Clausius-Clapeyron equation, given by
\begin{eqnarray}
\frac{d P}{dT}=\frac{\Delta S}{\Delta V},
\label{cc}
\end{eqnarray}
where $\Delta S$ and $\Delta V$ are the difference between the entropy and volume of the constituent in two different phases.

Note that the derivation of the above relation comes from the definition of the Gibbs potential $G=U-TS+PV$  ($U$ is the internal energy), where it is assumed that the first order derivatives of $G$ with respect to the intrinsic variables ($P$ and $T$) are discontinuous and it comes out that their specific values in two phases give the slope of the coexistence curve at the phase transition point.  

\section{Phase transition in AdS black holes}
As evident in the last section, the gibbsian approach is an effective tool for discussing phase transitions within the realm of conventional thermodynamics. Also, as we have already highlighted in our works \cite{bss}-\cite{modak}, this approach is equally powerful to study black hole phase transitions. Here we further elaborate and extend such ideas and show that this formalism illuminates issues concerning the stability/ instability of AdS black holes. As a byproduct of the analysis, some key results obtained by Hawking and Page \cite{hp1} are reproduced as well as generalized in a simple manner. 

Before we start giving our algebraic details, let us mention that throughout this letter, the various black hole parameters $M,~Q,~J,~S,~T$ should be interpreted as $\frac{M}{l},~\frac{Q}{l},~\frac{J}{l},~\frac{S}{l^{2}},~Tl$ respectively, where the symbols have their standard meanings. Also, with this convention, the parameter $ l $  no longer appears in any of the equations \cite{bgr}. 

We start by considering the R-N AdS black hole. The Gibbs energy for RN-AdS black hole is defined as $G = M-TS-\Phi Q$ where the last term is the analog of $PV$ term (with asign difference) in conventional systems. The expressions for mass ($M$), temperature ($T$), charge ($Q$) and the entropy ($ S $) are respectively given by \cite{bgr}
\begin{eqnarray}
M &=& \frac{\sqrt{S}[\pi(1+\Phi^2)+S]}{2\pi^{\frac{3}{2}}} \label{mr}\\
T &=& \frac{\pi(1-\Phi^2)+3S}{4\pi^{3/2}\sqrt{S}},\label{temprn}\\
Q &=& \Phi\sqrt{\frac{S}{\pi}}\label{q}\\
S &=& \pi r_{+}^{2}\label{s}.
\end{eqnarray}
where $ \Phi $ is the electric potential and $ r_+ $ is the radius of the outer event horizon.
Using these expressions one can easily write $G$ as a function of ($S,~\Phi$),  
\begin{eqnarray}
G &=& \frac{\sqrt{S}}{4\pi^{3/2}}\left(\pi(1-\Phi^2)-S\right), 
\label{grn1}
\end{eqnarray}

Considering a grand canonical ensemble (fixed $\Phi$) it is now straightforward to find the specific heat at constant potential ($C_{\Phi}$), which is the analog of $C_{P}$ (specific heat at constant pressure) in conventional systems. This is found to be
\begin{eqnarray}
C_{\Phi} =T\left(\frac{\partial S}{\partial T}\right)_{\Phi}=2S\frac{\pi(1-\Phi^2)+3S}{3S-\pi(1-\Phi^2)}, \label{sphrn}
\end{eqnarray}

With this machinery we are in a position to describe phase transitions in AdS black holes within the scope of standard thermodynamics. As a first step we plot Gibbs free energy, entropy and specific heat against temperature in figure 1. All these plots have a common feature; they  exist only when the temperature of the system is greater than the critical temperature ($T\ge T_0$). At $T_0$ they all show nontrivial behavior and, as we subsequently prove, there is a well defined phase transition at this temperature.

First, consider the $G-T$ plot. It has two wings which are joined at temperature $T_0$. At the upper wing, because of the positive definite values of $G$, this system falls in an unstable phase (Phase-1). At $T_0$ Gibbs free energy is maximum which implies that the system is most unstable at this point. Therefore it cannot stay at $T_0$ for long and eventually passes to the locally stable phase (Phase-2) by minimizing its free energy. This is achieved by following the other (lower) wing. In addition to that at $T=T_0$ the entropy-temperature plot is continuous and also changes its direction. The slope of this plot is negative for entropy lower than a critical value ($S_0$) while it is positive for all values higher than $S_0$. Entropy being proportional to the mass of a black hole, phase-1 in $ S-T $ plot corresponds to the lower mass (unstable) black holes while phase-2 belongs to black holes with higher mass (locally stable). This behavior is further exemplified in the plot of specific heat with temperature. The negative slope in $ S-T $ results in a negative heat capacity in phase-1. As the system approaches the critical point,  $ C_{\Phi} $ diverges. Exactly at $ T_0 $ it flips from negative infinity to positive infinity \cite{hp7}. Through this phase transition the system emerges into a locally stable phase (phase-2) where the specific heat becomes positive. For $ T<T_0 $ black holes do not exist and what remains is nothing but the thermal radiation in pure AdS space. 

Furthermore from the $G-T$ plot we find that $G$ changes its sign at temperature $T=T_1$. The free energy is thus always positive for  $ T_0<T< T_1$. For $ T>T_1 $ the system is globally stable. At $T_1$ and we have the well known Hawking-Page transition \cite{hp9}-\cite{hp10}. 

We now give algebraic details. As a first step, the minimum temperature $ T_0 $ is calculated for the RN AdS space. This is found from the temperature at which $ C_{\Phi} $  diverges (see fig.1). From (\ref{sphrn}) this occurs at the critical entropy, 
\begin{equation}
S_{0}=\frac{\pi}{3}(1-\Phi^2).
\label{sc}
\end{equation}
Substituting this value in (\ref{temprn}) we find the critical temperature
\begin{equation}
T_{0}=\frac{1}{2\pi}\sqrt{3(1-\Phi^2)}.
\label{tc}
\end{equation} 
In the charge less limit ($\Phi=0$), this result goes over to,
 \begin{equation}
 T_0=\frac{\sqrt{3}}{2\pi}
 \end{equation}
 This reproduce the result for the Schwarzschild AdS black hole found earlier in \cite{hp1}{\footnote{Note that $l$ does not appear since as explained before, it has been appropriately scaled out.}}. 

\begin{figure}[h]
\centering
\includegraphics[angle=0,width=12cm,keepaspectratio]{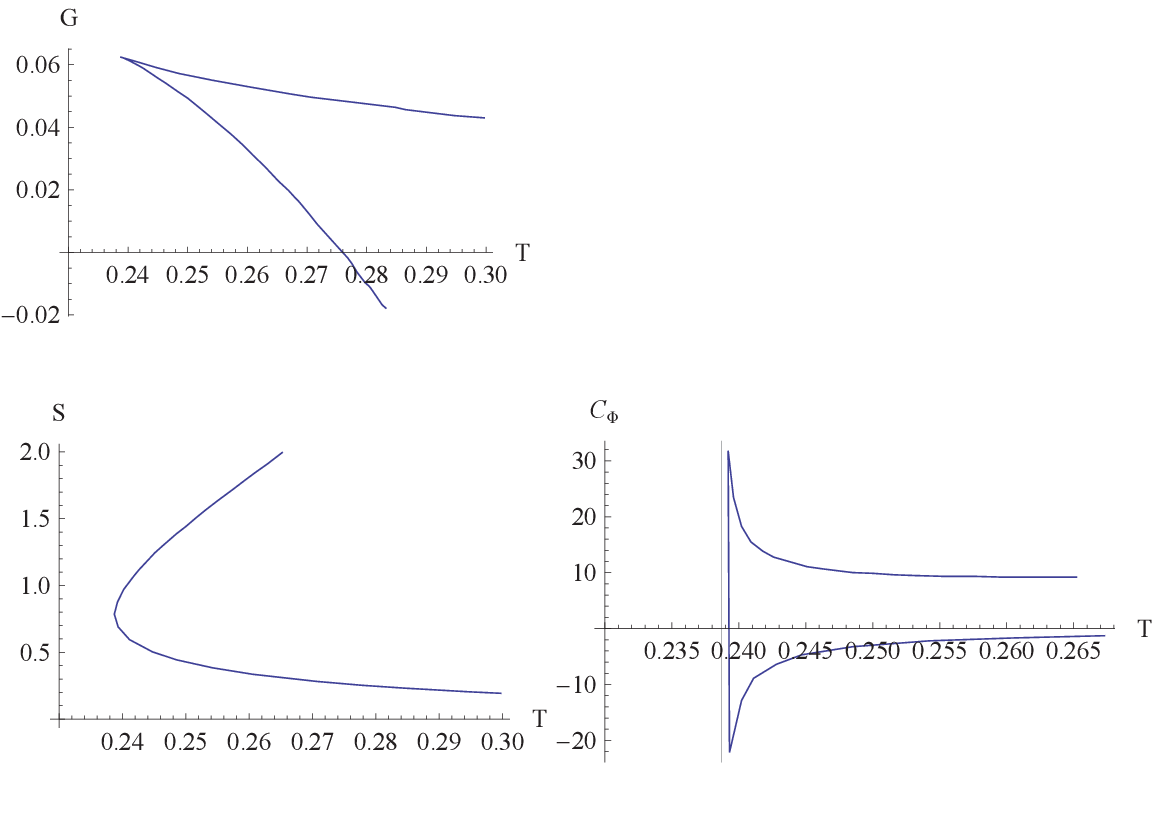}
\caption[]{\it Gibbs free energy ($G$), entropy ($S$) andspecific heat ($C_{\Phi}$) plot for R-N AdS black hole with respect to temperature ($T$) for fixed $\Phi=0.5$. For detailed discussion see text.}
\label{figure 2a}
\end{figure} 
Observe that (\ref{tc}) also follows by following the original analysis of \cite{hp1}. Following this approach we first calculate the minimum temperature $T_0$ for the RN-AdS space-time. One can construct thermal states in RN-AdS space time by periodically identifying the imaginary time coordinate $ \tau $ with period 
\begin{equation}
\beta=\frac{4\pi^{3/2}\sqrt{S}}{\pi(1-\Phi^2)+3S}
\end{equation}
which is the inverse Hawking temperature ($T$). For the maximum value of $\beta$ (i.e. minimum value of $T$) at we set $\left[ \partial\beta/\partial r_{+}\right] _{\Phi}=0$. From this condition and by using (\ref{s}) we find $S=\frac{\pi}{3}(1-\Phi^2)$. Substituting this into (\ref{temprn}) we find the corresponding minimum temperature for RN-AdS space time to be identical with (\ref{tc}).

The critical mass separating large and small black holes is now deduced. By substituting (\ref{sc}) in (\ref{mr}) we obtain,
\begin{equation}
M_{0}=\frac{\sqrt{1-\Phi^2}(2+\Phi^2)}{3\sqrt{3}}.
\label{mc}
\end{equation}
In the charge less limit ($ \Phi=0 $) it yields,
\begin{equation}
M_0=\frac{2}{3\sqrt{3}}
\end{equation}
which again reproduces the result given in \cite{hp1}.
Finally, the value of  $T_{1}$ is obtained from (\ref{grn1}) and (\ref{temprn}),
\begin{equation}
T_1=\frac{\sqrt{1-\Phi^2}}{\pi}.
\label{t'}
\end{equation}
If we make $\Phi=0$, the temperature $T_1$ simplifies to,
\begin{equation}
T_1=\frac{1}{\pi}
\end{equation}
which is, once again, the result for the Schwarzschild AdS black hole \cite{hp1}.

For the Kerr-AdS black hole one can easily perform a similar analysis and describe phase transition for that case. The graphical analysis shown above for RN-AdS black hole physically remains unchanged for the Kerr-AdS case. In particular one can calculate the inverse temperature 
\begin{eqnarray}
\beta^{Kerr}=\frac{4\pi^{\frac{3}{2}}{[S(\pi+S)(\pi+S-S\Omega^2)]^{1/2}}}{\pi^2-2\pi S(\Omega^2-2)-3S^2(\Omega^2-1)}
\label{temp}
\end{eqnarray}
and the condition $\left[ \partial\beta^{Kerr}/\partial r_{+}\right] _{\Omega}=0$ gives 
\begin{eqnarray}
(\pi+S)^3(3S-\pi)-6S^2(\pi+S)^2\Omega^2+S^3(4\pi+3S)\Omega^4=0.
\label{betm}
\end{eqnarray}
By substituting the positive solution of this polynomial in (\ref{temp}) one finds the minimum temperature $T^{Kerr}_0$. Furthermore for the Kerr-AdS black hole the heat capacity diverges exactly where (\ref{betm}) holds \cite{modak} which corresponds to the minimum temperature ($T^{Kerr}_{0}$). In the irrotational limit $T^{Kerr}_0$ reproduces the known result of \cite{hp1}. Results for $M_0$ and $T_1$ can also be calculated by a similar procedure. These expressions are rather lengthy and hence omitted. Moreover there are no new insights that have not already been discussed in the RN-AdS example.

In the remaining part of this paper we shall analyze and classify the phase transition at temperature $T_0$ by exploiting Ehrenfest's scheme. Since Hawking-Page transition at temperature $T_1$ is already studied extensively in the literature we do not include this part in our paper.  

\section{Analytical check of Ehrenfest relations}
Note that the ($S-T$) graph (fig.1) shows entropy is a continuous at temperature $T_0$. Consequently a first order transition is ruled out. However, the infinite discontinuity in specific heat (see fig.1) strongly suggests the onset of a higher order (continuous) phase transition. Under this circumstance, Ehrenfest's equations are expected to play a role. The derivation of these equations demands the continuity of specific entropy ($S$), specific charge ($Q$) and specific angular momentum ($J$) at the critical point. As a result these relations are truly local and only valid infinitesimally close to the critical point. For a genuine second order phase transition both equations have to be satisfied \cite{stanley,zeman}. 

\subsection{R-N AdS black hole}
We start by considering the RN-AdS black hole. Ehrenfest's equations for this system are given by \cite{bgr} 
\begin{eqnarray}
&& -\left(\frac{\partial \Phi}{\partial T}\right)_{S} = \frac{C_{\Phi_2}-C_{\Phi_1}}{T Q(\alpha_2-\alpha_1)}
\label{ehf1}\\
&&-\left(\frac{\partial \Phi}{\partial T}\right)_{Q} = \frac{\alpha_{2}-\alpha_{1}}{k_{T_{2}}-k_{T_{1}}}
\label{ehf2}
\end{eqnarray} 
where, $\alpha =\frac{1}{Q}\left( \frac{\partial Q}{\partial T}\right) _{\Phi}$ is the analog of volume expansion coefficient and $k_T =\frac{1}{Q}\left( \frac{\partial Q}{\partial \Phi}\right) _{T}$ is the analog of isothermal compressibility. Their explicit expressions are given by \cite{bgr}
\begin{eqnarray}
Q\alpha &=& \frac{4\pi \Phi S}{3S-\pi(1-\Phi^{2})} \label{palpha}\\
Qk_T &=& \sqrt\frac{S}{\pi}\frac{3 \pi\Phi^{2}-\pi +3S}{3S-\pi(1-\Phi^{2})}
\label{pkappa}
\end{eqnarray}
The L.H.S. of both Ehrenfest's equations (\ref{ehf1}) and (\ref{ehf2}) are found to be identical at the critical point $S_0$. Using the defining relations (\ref{temprn},\ref{q}) the slope of the coexistence curve at the phase transition point is found to be \cite{bgr}, 
\begin{equation}
-\left[ \left(\frac{\partial \Phi}{\partial T}\right)_{S}\right] _{S=S_0}=-\left[ \left(\frac{\partial \Phi}{\partial T}\right)_{Q}\right]_{S=S_0} =\frac{2\surd\pi \surd S_{0}}{\Phi}.
\label{eh1}
\end{equation} 

In order to calculate the right hand sides, $\Phi$ must be treated as a constant (this is analogous to fixing pressure while performing an experiment). This would help us to re-express $C_\Phi$ (\ref{sphrn}), $Q\alpha$ (\ref{palpha}), and $Qk_T$ (\ref{pkappa}), which have functional forms $\frac{f(S)}{g(S)}$,  $\frac{h(S)}{g(S)}$  and  $\frac{k(S)}{g(S)}$ respectively,  infinitesimally close to the critical point ($S_0$). Note that they all have the same denominator which satisfies the relation $g(S_0)= 3S_0-\pi(1-\Phi^{2})=0$. This observation is crucial in the ensuing analysis.  

The expressions of $C_{\Phi}$, $Q\alpha$ and $Qk_T$ in the two phases ($ i=1,2 $) are respectively given by ${C_{\Phi}}\big|_{S_i}=C_{\Phi_i}$, ${Q\alpha}\big|_{S_i}=Q\alpha_i$ and $Qk_T\big|_{S_i}=Qk_{T_i}$. To obtain the R.H.S. of (\ref{ehf1}) we first simplify it's numerator:
\begin{eqnarray}
C_{\Phi_{2}}-C_{\Phi_{1}}=\frac{f(S_2)}{g(S_2)}-\frac{f(S_1)}{g(S_1)}
\end{eqnarray} 

Taking the points close to the critical point we may set  $f(S_2)=f(S_1)=f(S_0)$ since $f(S)$ is well behaved. However since $ g(S_0)=0 $ we do not set $g(S_2)=g(S_1)=g(S_0)$. Thus 
\begin{eqnarray}
C_{\Phi_{2}}-C_{\Phi_{1}}=f(S_0)\left(\frac{1}{g(S_2)}-\frac{1}{g(S_1)}\right).
\end{eqnarray} 
Following this logic one derives,
\begin{eqnarray}
\frac{C_{\Phi_2}-C_{\Phi_1}}{T_0 Q(\alpha_2-\alpha_1)}=\frac{f(S_0)}{T_0 h(S_0)}=\frac{2\surd\pi \surd S_{0}}{\Phi}
\label{eh2}
\end{eqnarray}
and, similarly,
\begin{eqnarray}
\frac{Q(\alpha_{2}-\alpha_{1})}{Q(k_{T_2}-k_{T_1})}=\frac{h(S_0)}{k(S_0)}=\frac{2\surd\pi \surd S_{0}}{\Phi}.
\label{eh3} 
\end{eqnarray}

Remarkably we find that the divergence in $C_{\Phi}$ is canceled with that of $\alpha$ in the first equation and the same is true for the case of $\alpha$ and $k_T$ in the second equation. From (\ref{eh1},\ref{eh2},\ref{eh3}) the validity of the Ehrenfest's equations is established. Hence this phase transition in RN-AdS black hole is a genuine second order transition.

\subsection{Kerr AdS black hole}
Ehrenfest's set of equations for Kerr-AdS black hole are given by \cite{bss,modak}
\begin{eqnarray}
-\left(\frac{\partial \Omega}{\partial T}\right)_{S} &=& \frac{C_{\Omega_2}-C_{\Omega_1}}{TJ(\alpha_2-\alpha_1)},\label{ehrk1}\\
-\left(\frac{\partial \Omega}{\partial T}\right)_{J} &=& \frac{\alpha_{2}-\alpha_{1}}{k_{T_{2}}-k_{T_{1}}}.\label{ehrk2}
\end{eqnarray}
The expressions for specific heat ($C_{\Omega}$), analog of the volume expansion coefficient ($\alpha$) and compressibility ($k_T$) are all provided in \cite{modak}. Once again, they all have the same denominator (like the corresponding case for RN-AdS). Considering the explicit expressions given in \cite{modak} and using the same techniques we find that both sides of (\ref{ehrk1}) and (\ref{ehrk2}) lead to an identical result, given by, 
\begin{equation}
l.h.s=r.h.s=\frac{4\pi^{\frac{3}{2}}(\pi+S_0-S_0\Omega^2)^{\frac{3}{2}}(\pi+S_0)^{\frac{1}{2}}}{\sqrt{S_0}\Omega[3(\pi+S_0)^2-S_0\Omega^2(2\pi+3S_0)]} \label{kerr}.
\end{equation}
 This shows that the phase transition for Kerr-AdS black hole is also second order. In our earlier work \cite{modak} the above equality was shown numerically for various values of $ \Omega $. Thus the analytical result (\ref{kerr}) is compatible with our previous finding \cite{modak}.




\section{Conclusions}
In the present paper we established that the concepts of thermodynamics which describe the well known liquid-vapour phase transitions are equally capable to understand black hole phase transitions. This is true despite the fact that black hole entropy is {\it non-extensive}. In fact black holes play the role of a theorist's laboratory to check the ideas based on Clausius-Clapeyron-Ehrenfest schemes. 

This paper also completes a formulation to discuss phase transition in AdS black holes following a conventional thermodynamical approach. This further clarify the interpretation of black holes as thermodynamic objects. Considering a grand canonical ensemble the critical temperature and mass of the charged (Reissner-Nordstrom) AdS black holes undergoing a phase transition were calculated. The phase transition discussed in this paper was defined by the discontinuity of specific heat, volume expansion coefficient and compressibility. The two phases were identified with black holes having `smaller' and `larger' mass/ horizon radius. The branch with smaller mass has negative specific heat and positive free energy, while, the other (larger mass) branch has positive specific heat and positive free energy (less than the free energy of smaller mass black holes). The phase transition occurred from a lower mass black hole to a higher mass black hole. Thus it was a transition between an unstable phase to a locally stable phase. Although various relevant thermodynamic variables diverged at the critical point, we were able to devise an {\it analytical approach} for checking the Ehrenfest's relations appropriate for a second order transition. The exact validity of these relations confirmed the second order nature of the phase transition. Similar results for the rotating (Kerr) AdS black holes were outlined. These were found to be compatible with our earlier work \cite{modak} which employed a numerical approach. As a future work we plan to build a proper microscopic description of black hole phase phase transitions which is still not known.

\section*{ Acknowledgement:}
 S. K. M and D.R like to thank the Council of Scientific and Industrial Research (C. S. I. R), Government of India, for financial help.

  
\end{document}